\def\rg{$r_{\rm g}$}
\def\phpspsqcm{ph\thinspace s$^{-1}$\thinspace cm$^{-2}$}
\def\cm{{\rm\thinspace cm}}

\def\erg{{\rm\thinspace erg}}
\def\eV{{\rm\thinspace eV}}

\def\keV{{\rm\thinspace keV}}

\def\Mpc{{\rm\thinspace Mpc}}
\def\Msun{\hbox{$\rm\thinspace M_{\odot}$}}
\def\Zsun{\hbox{$\rm\thinspace Z_{\odot}$}}
\def\pc{{\rm\thinspace pc}}

\def\s{{\rm\thinspace s}}

\def\Hz{{\rm\thinspace Hz}}

\def\ergpcmsqps{\hbox{$\erg\cm^{-2}\s^{-1}\,$}}

\def\ergps{\hbox{$\erg\s^{-1}\,$}}

\def\pcmsq{\hbox{$\cm^{-2}\,$}}

\def\spose#1{\hbox to 0pt{#1\hss}}
\def\approxlt{\mathrel{\spose{\lower 3pt\hbox{$\sim$}}
        \raise 2.0pt\hbox{$<$}}}
\def\approxgt{\mathrel{\spose{\lower 3pt\hbox{$\sim$}}
        \raise 2.0pt\hbox{$>$}}}
\def\rg{r_{\rm g}}

\def\ergcmps{erg\thinspace cm\thinspace s$^{-1}$}
\def\Ms{{\it M}$_\odot$}

\documentclass{mn2e}
\usepackage{times}

\input{psfig.sty}

\newif\ifAMStwofonts

\title[An X-ray variability study of NGC 4395]
        {Evidence for an intermediate mass black hole and a multi-zone warm absorber in NGC 4395}
\author[D.~C.~Shih et al.]
        {D.~C.~Shih,$^{1,2}$ K.~Iwasawa$^1$ and A.~C.~Fabian$^1$
\\$1$ Institute of Astronomy, Madingley Road, Cambridge, CB3 0HA
\\$2$ Department of Physics, Princeton University, Princeton, NJ 08544
}

\date{}

\begin{document}

\maketitle

\label{firstpage}

\begin{abstract}
We report on the results of an analysis in the X-ray band of a 
recent long \textit{ASCA} observation of NGC 4395, the most variable 
low-luminosity AGN known. A relativistically-broadened iron line at 
$\sim$~6.4~keV is clearly resolved in the time-averaged spectrum,
with an equivalent width of $310^{+70}_{-90}$~eV. 
Time-resolved spectral analysis of the heavily absorbed soft X-ray band
confirms the existence of a variable, multi-zone warm absorber in this 
source, as proposed in a previous analysis of a shorter \textit{ASCA} 
observation. The light curve of the source is wildly
variable on timescales of hours or less, and a factor 
of nearly 10 change in count-rate was recorded in a 
period of less than 2000~s. The long observation and variability
of the source allowed the power density spectrum (PDS) to be constructed
to an unprecedented level of detail. There is evidence for a
break in the PDS from a slope of $\alpha \sim 1$ to $\alpha \sim 1.8$ 
at a frequency of around $3 \times 10^{-4}$~Hz. The central black hole mass
of NGC 4395 is estimated to be approximately $10^4$--$10^5$ M$_{\odot}$ using
the break in the PDS, a result consistent with previous analyses using optical
and kinematical techniques.
\end{abstract}

\begin{keywords}
galaxies: individual: NGC 4395 -- galaxies: Seyfert -- X-rays: galaxies.
\vspace*{1cm}
\end{keywords}

\section[Introduction]
{Introduction}

NGC 4395 is a nearby ($d \sim$ 2.6 \Mpc) dwarf galaxy containing an active nucleus. 
The nucleus harbors a small black hole of mass less than
$8\times 10^5$ \Ms~\cite{Filippenko2002}. The optical emission-line
spectrum of the nucleus is dominated by non-stellar emission,
reminiscent of normal Seyfert galaxies but with a much lower
luminosity \cite{Ho97}.  Indeed, NGC 4395 is one of the
lowest-luminosity AGN, with an absorption-corrected 2--10~keV luminosity of $4\times
10^{39}$ \ergps. However, in contrast to other low luminosity
active nuclei, which are usually found at the centers of galaxies having
large bulges and which exhibit weak X-ray variability \cite{Ptak98},
NGC 4395 has no significant bulge and shows strong X-ray variability. Interestingly,
it appears to lie on the anti-correlation between X-ray variability amplitude and 
luminosity found in higher luminosity Seyfert nuclei \cite{Nandra97a}.
In terms of a possible link between black hole mass and X-ray variability,
NGC 4395 is therefore of great interest, given the fact that it is one
of the few objects containing a small black hole of well-determined mass.

It is a well-established fact that the X-ray light curves of active galaxies show
a red-noise characteristic, and approximating the power spectrum with a power law
typically gives slopes of 1.5--2 on time scales of hours
to days. Evidence for a flattening of the power-law slope on longer time scales,
similar to that found in Galactic Black Hole Candidates (GBHC) but scaled to much shorter
time scales, has been found in some active galaxies. Being a low-mass
black hole AGN, NGC 4395 provides us with an opportunity to examine the interesting 
part of the power spectrum with a relatively short monitoring time compared with
its higher luminosity counterparts.

In addition to extreme X-ray flux variability, NGC 4395 is also known to exhibit 
interesting spectral variability as well. The source was previously observed
by \textit{ASCA} in 1998, and an analysis by Iwasawa et al.\ (2000) found
the time-averaged X-ray spectrum to be well-described 
by a power-law continuum with a photon index of 1.5--2, 
with heavy absorption below $\sim$ 3~keV and
marginal evidence for iron line emission at $\sim$ 6.4~keV. During
a large flare in the 1998 observation, the spectrum was observed 
to vary dramatically below 3~keV, with the absorption appearing
to lessen considerably between 1--2~keV. A single-zone
warm absorber was found to give a physically inconsistent description
of the active/quiescent states, and a two-zone
warm absorber consisting of a constant and a variable component
was proposed to explain the observed spectral variability.

In this paper, we present the results of an analysis of a week-long 
\textit{ASCA} observation of NGC 4395, focusing on the temporal and spectral 
variability of the X-ray source. Since the previous observation
was only a half-day in duration, the conclusions reached in
this analysis represent a significant improvement in our understanding
of this source.

\section{Observation and Data Reduction}
\label{sec_obs_and_datared}

\begin{figure*}
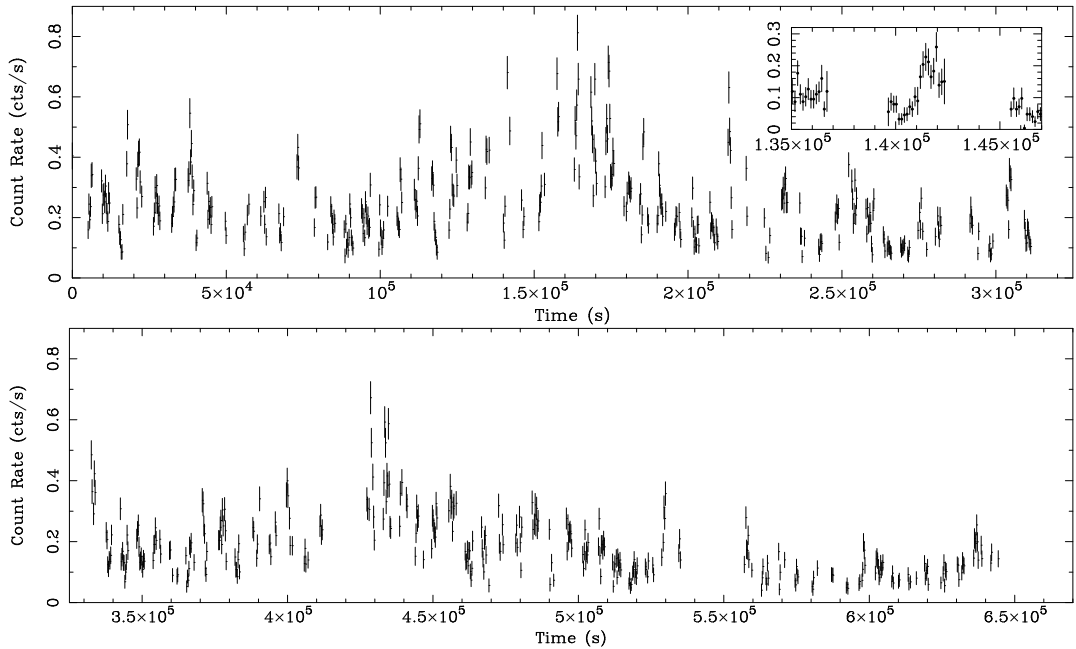

\centerline{\psfig{figure=lc.sis+gis.256tb.1.2-10kev.ps.t0-325000.2,width=0.8\textwidth,angle=270}}
\centerline{\psfig{figure=lc.sis+gis.256tb.1.2-10kev.ps.t325000-650000,width=0.8\textwidth,angle=270}}
\caption{The 1.2--10~keV band SIS+GIS background-subtracted 
light curve with 256-s time bins. 
The inset shows (for the SIS0 detector only and with 128-s bins) a 
particularly fast X-ray variation, with the count rate increasing by 
a factor of nearly 10 in $\sim 1500$~s.}%
\label{fig_lc_ngc4395}
\end{figure*}

NGC 4395 was observed by \textit{ASCA} from 2001 May 25 to 2001 June
1, for a total exposure time of 640 ks. Data reduction was carried out
using the method of Iwasawa et al.\ (2000).  Following GTI filtering,
the good exposure time was approximately 300 ks for the SIS and 280 ks
for the GIS. The mean count rates for the SIS/GIS (using the
large-aperture region -- see below) were $0.075/0.079$ cts/s in the
2--10~keV band and $0.038/0.028$ cts/s in the 1--2~keV band. The
observed 2--10~keV flux averaged over the whole observation is
$4.7\times 10^{-12}$\ergpcmsqps, very similar to that obtained from
the previous short \textit{ASCA} observation, while the source is highly
variable. The time-averaged 1--2~keV flux is $3.1\times 10^{-13}$\ergpcmsqps,
somewhat higher than the corresponding flux of $2.6\times 10^{-13}$\ergpcmsqps
of the previous \textit{ASCA} observation.



Several circumnuclear X-ray sources are known to exist within 10
arcmin of the nucleus of NGC 4395 \cite{Moran99,Lira99,Iwasawa2000}.
In this observation, we could only detect the brightest of these
background sources, corresponding to source E of Moran et al.\ (1999),
a very soft source located $\sim 2.8$ arcmin away from the nucleus,
with X-ray emission peaking below 1 keV. We have checked that the mean
fluxes from source E and the nucleus in various energy bands are in
agreement with the previous \textit{ASCA} observation within 15 per
cent, and that source E does not show significant variability.
Therefore, we have chosen to use in this analysis the source regions
defined by Iwasawa et al.\ (2000) in their analysis of the previous
observation.  We briefly describe here the definition of these
regions. Since the flux of source E is negligible relative to the
nucleus above $\sim 2$~keV, we use photons collected from a
large-aperture (6 arcmin diameter) circular region around the nucleus
(but excluding a small 1 arcmin circular region around source E) for
analysis above 2~keV. Below 2~keV, the contamination due to source E
becomes non-negligible owing to the broad point spread function of the
\textit{ASCA} X-ray telescope. In fact it may be as high as 70\% in
the large-aperture region. To minimize this contamination we use
instead a small-aperture (3 arcmin diameter) region around the nucleus
for any analysis requiring data below 2~keV (e.g.\ analysis of the
warm absorber) so that the contamination from the source E in the 1--2
keV band should be less than 10 per cent (see Iwasawa et al. 2000).

\section{Timing Analysis}

\subsection{X-ray Light Curves}
\label{sec_RMS}

The strong X-ray variability of NGC 4395 suggested by the previous short \textit{ASCA}
observation is confirmed in this long observation, as can be seen in the light curve 
shown in Fig.\ \ref{fig_lc_ngc4395}. The light curve is the sum of the SIS and GIS 
light curves and includes background subtraction. Each point represents a full 256-s of
good exposure time. The source is seen to vary wildly on timescales of hours or less. 
In several instances, the count rate was observed to change by nearly an 
order of magnitude on time scales of $\sim 2000$~s; the most dramatic example
is shown in the inset of Fig.\ \ref{fig_lc_ngc4395}.

\begin{figure}
\centerline{\psfig{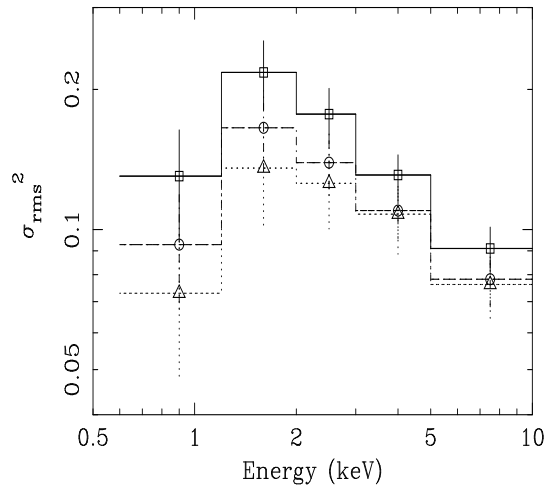}}
\caption{The normalized RMS variance spectrum on 12 ks (squares), 23 ks (circles), and 46 ks (triangles) timescales.
The peak in RMS variance in the 1--2~keV band is most likely due to a variable warm absorber, a hypothesis
that we shall explore in Section \ref{sec_multizone_warmabs}.}
\label{fig_rms}
\end{figure}

The energy dependence of the normalized excess variance $\sigma_{\rm
  rms}^2$ (see Nandra et al 1997 for the definition) of the light
curve is plotted in Fig.\ \ref{fig_rms}. On a wide range of
timescales, the strongest variability is found in the 1--2~keV band.
As mentioned above, any contamination from source E has little effect
on this excess variability in this energy range. The variability of
the warm absorber, and not of the primary continuum, is likely to be
the main contributor to the peak in the RMS variance spectrum. Indeed,
in the context of the multi-zone warm absorber model to be discussed
in detail in Section \ref{sec_multizone_warmabs}, we see that the
variability in the 1--2~keV band can be attributed to rapid changes in
a highly ionized, dense absorber located close to the source.

\subsection{The Power Density Spectrum}
\label{sec_PDS}

The power density spectrum (PDS) of NGC 4395 was computed for the 1.2--10~keV band 
light curve using data from all four detectors following the method of Hayashida et al.\ (1998). 
The one difference with the method of Hayashida et al.\ (1998) was that geometric rather
than arithmetic averaging was employed to smooth the PDS. Smoothing the logarithm of the PDS 
instead of the PDS itself tends to give a less biased estimate of the PDS
when the number of points in each frequency bin is small \cite{Papadakis93}.
5760-s bins were used to obtain the low frequency PDS and 128-s bins to obtain the high frequency PDS. 
These binsizes were also chosen so as to avoid frequencies around $10^{-4}$ \Hz, where 
contamination from the orbital period of the spacecraft produces a spurious peak in the PDS. 
Segments with less than four time bins were not used in the calculation 
in an attempt to limit the effect of red-noise leak. The low frequency PDS was smoothed 
with $N=5$ and the high frequency PDS with $N=25$.      

\begin{figure}
\centerline{\psfig{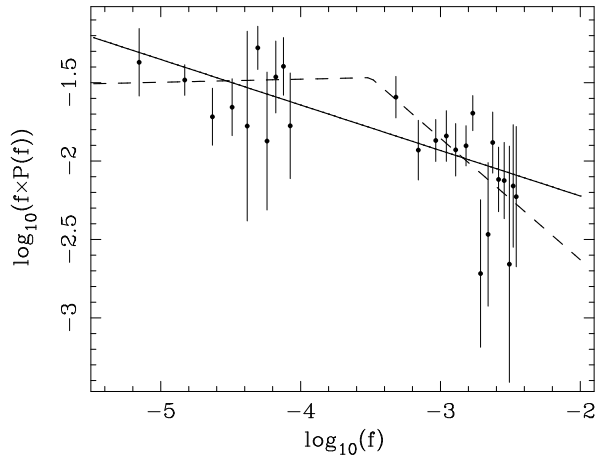}}
\caption{The normalized power density spectrum (NPDS) of NGC 4395 derived
from the 1.2--10~keV band light curve, plotted as frequency times power.
The solid line shows the best-fit single power law to the entire PDS
($\alpha=1.29$), while the dashed line shows the best-fit broken power law
($\alpha_1=0.98$, $\alpha_2=1.78$, $f_{\rm b}= 3.2\times10^{-4}$~\Hz).}
\label{fig_powerspectrum}
\end{figure}

The result of the calculation using the 1.2--10~keV light curve is shown in Fig.\ \ref{fig_powerspectrum}.
We report a possible detection of a break in the power spectrum. A single power-law with best-fit
slope $\alpha = 1.29 \pm 0.04$ gives a relatively poor description of the PDS 
($\chi^2 = 37.6$ for 24 degrees of freedom). A broken power-law with
$\alpha_1 = 0.98 \pm 0.18$ ($f < f_{\rm b}$), $\alpha_2 = 1.78 \pm 0.19$ ($f \geq f_{\rm b}$), and 
break frequency $f_{\rm b} = 3.2^{+3.2}_{-1.6}\times 10^{-4}$ \Hz~gives an improved fit
($\chi^2 = 27.9$ for 22 degrees of freedom). The improvement is significant at a 96\% confidence level
according to the standard F-test. We note, however, that the F-test is a crude measure of significance 
and must be interpreted with caution. In particular, it does not take into account systematic biases 
such as aliasing and red-noise leak which could cause the spurious detection of a break.

The PDS was also calculated using the 2--10~keV light curve to check whether variability of the warm 
absorber below $\sim$ 2~keV could have a significant effect on the parameters of the 
measured power spectrum. The broken power law slopes, and more importantly, the \textit{location
of the break} measured from the 2--10~keV band PDS agrees within errors
with those quoted above. It was noted in Section \ref{sec_RMS} that the RMS variance spectrum
peaks in the 1--2~keV band, most likely due to changes in the warm absorber and not the source 
itself. However, the fact that the break in the PDS is present and has the same frequency (within errors)
in both the 1.2--10~keV band and the 2--10~keV band suggests that the break is an intrinsic feature of
the source. Indeed, since the shape of the PDS is similar between the bands,
the 1.2--10~keV band PDS seems to provide a measure of the intrinsic source variability. The additional
effect of the variable warm absorber is undetectable in the PDS owing to its limited 
bandwidth (1.2--2~keV), the high intrinsic variability of the source, and the large uncertainties of
the measured PDS.

The shape of the PDS is remarkably similar to those of
Galactic Black Hole Candidates (GBHC) in the low state. The PDS of Cygnus X-1, 
among the most famous and well-studied GBHC, has been measured to very high accuracy over nearly five 
decades in frequency in a number of observations \cite{Gilfanov99,Nowak99,Belloni90}. It too
exhibits a spectral steepening from $\alpha \approx 1$ to $\alpha \approx $ 1.5--1.7 as one increases
in frequency, while at lower frequencies the PDS hits a ``knee'' and becomes nearly flat. This
three-component behavior has also been observed in several other GBHC (e.g.\ Miyamoto 1992). While
a longer observation is needed to test for the presence of a flat component at very low frequencies in 
the PDS of NGC 4395, the similarity of the two higher-frequency components to the power spectra of GBHC 
suggests that the physical processes that give rise to X-ray variability in
GBHC are similar to those of NGC 4395. A better understanding of the X-ray variability of the former, enabled
by their much higher signal-to-noise and shorter variability time scales,  might therefore lead to 
improved knowledge of NGC 4395. 

As a preliminary attempt at a comparison, we can use the striking 
similarity between the power spectra of NGC 4395 and Cyg X-1 in 
order to obtain an estimate of the black hole mass of the former, following
the scaling method of Hayashida et al.\ (1998). We shall describe 
the technique and results in detail in Section \ref{sec_bhmass}.
Here, we merely note that in the analysis of the previous short \textit{ASCA} 
observation by Iwasawa et al.\ (2000), the mass
of NGC 4395 was estimated using the PDS to be about $3 \times 10^4 \Msun$. 
The PDS obtained in that analysis, however,
consisted of only a handful of points, and so neither the break frequency 
nor the slope of the PDS could be measured. Instead,
the authors used a template PDS taken from NGC 4051, which breaks from 
$\alpha = -0.28$ to $\alpha = 2.01$, and scaled it
to the PDS of NGC 4395 with the break frequency as a free parameter. 
With the much more detailed power spectrum 
from the current observation, we are actually able to both measure the break frequency and the 
spectral slopes, and we see that in fact the NGC 4051 PDS is a very poor template for NGC 4395. 

\section{Spectral Analysis}

\subsection{Time-Averaged Spectrum: Warm Absorber and Iron K$\alpha$ Line}
\label{sec_tavg}

\begin{figure}
\centerline{\psfig{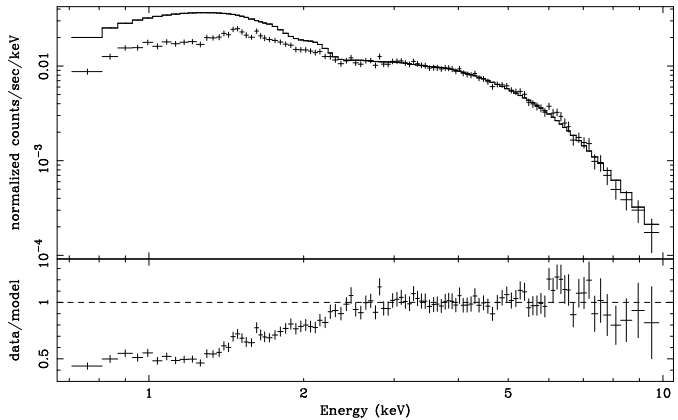}}
\caption{The time-averaged SIS spectrum, along with the best-fit power law 
to the 3--10~keV band. The ratio of data to model shown in the bottom panel illustrates the heavy absorption below 3~keV,
as well as the residuals at $\sim$ 6~keV due to iron-line emission. A moderately 
strong (EW $\sim$ 300~eV) broad iron line is detected in the present observation at the $99\%$ confidence level.}
\label{fig_ngc4395_tavgedspec}
\end{figure}

The X-ray spectrum of NGC 4395 is known to be heavily absorbed 
by what is thought to be ionized material along
the line of sight (the so-called warm absorber). With 
non-negligible absorption extending up to or even beyond 3~keV,
the likely presence of a broad iron line around 5--7~keV, 
a reflection hump from the disk starting at $\sim$ 10~keV,
and an iron K edge around 8~keV due to the warm absorber, 
obtaining a reliable value for the
continuum photon index is a highly non-trivial task. 
Fig.\ \ref{fig_ngc4395_tavgedspec} shows the 
result of a preliminary power-law fit to the SIS in 
the 3--10~keV continuum band, extrapolated to lower energies. 
The residuals at $\sim$ 6~keV are highly suggestive of 
iron-line emission. The extrapolation to lower energies reveals 
the tell-tale signature of warm absorption: a blend of 
absorption lines and edges due to partially-ionized, obscuring material. 

The best-fit photon index for this simple power-law fit was unusually
hard, $\Gamma \approx 1$.  A hard photon index ($\Gamma \approx 0.6$)
was also recently reported by Moran et al. (2001) during a recent
17~ks \textit{Chandra} observation of NGC 4395. The discrepancy between
the photon index measured in the present analysis and that of 
Moran et al. (2001) might be explained by the effects of photon pile-up.
The same Chandra data analyzed by Moran et al. (2001)
were found to be affected by the CCD pile-up \cite{Davis2001}, 
despite the attempt of reducing it by using faster readout. On the other
hand, pile-up should be completely neglible in the \textit{ASCA} data of
the present analysis. The point spread function of \textit{ASCA} is
far broader than that of \textit{Chandra}, and there is correspondingly very little
chance in the \textit{ASCA} data of more than two photons falling into a single pixel within
a single readout for a faint source such as NGC 4395.
Two important effects of the pile-up are: 1) the
light curve has reduced flux peaks where the pile-up occurs; and 2)
the spectrum is distorted to have a harder slope than is actually present.
The second effect might have resulted in the very small photon
index measured by Moran et al. (2001). The range of X-ray flux
variations in NGC 4395 is quite large: from very low count rates where
the pile-up is negligible to high count rates where significant
pile-up occurs. The high-flux state is where the spectral softening
due to 1--2 keV flux increase has been found in the two \textit{ASCA}
observations (e.g., see Fig. 6). The spectral distortion due to the
pile-up would reduce this spectral softening for high flux state if it
was not corrected, affecting a spectral variability study. In any case, it is
extremely unlikely that these low photon indices are associated with
the intrinsic source spectrum, given the standard mechanism of
continuum production via inverse-Compton scattering of thermal disk
photons off a hot, optically-thin corona \cite{Moran2001}.  A more
plausible explanation is that the source has an intrinsically steep
spectrum, that is then modified, perhaps by obscuring material along
the line of sight, to produce a flat spectrum. Moran et al. (2001)
suggest that this obscuring material could be due to neutral ``partial
covering'' components. As they note in their paper, however, this
partial-covering model predicts a strong Fe K edge near 7~keV that is
not observed. Below, we will consider another possible explanation for
the flatness of the continuum, namely that the (multi-zone) warm
absorber is so heavily ionized, that absorption above 3~keV flattens
the observed spectrum.

To investigate the residuals at 6~keV, a relativistically broadened
6.4~keV iron line was added to the spectrum using the
{\tt diskline} model in {\sc xspec} \cite{Fabian89}. The free 
parameters were the inclination 
angle $\theta_0$ and emissivity index $\alpha$ of the accretion disk,
and the total line flux $F_{{\rm K}\alpha}$. The inner and outer radii were frozen at 
6 $\rg$ and 100 $\rg$. In addition, cold absorption was included (with
the column density as a free parameter) in order to model the
effect of the warm absorber above 2~keV. Since we only expect to see the
smooth tails of the ionized warm absorber edges above 2~keV, these can be approximated
phenomenologically by the {\tt wabs} neutral absorption model in {\sc xspec}. 
Finally, the effects of neutral reflection by the accretion disk were taken into account by
using {\tt pexrav} instead of a simple power-law for the underlying continuum. The power-law
index $\Gamma$ and overall normalization were left free to vary. The remaining {\tt pexrav} parameters
were frozen at reasonable values: the high-energy cut-off, reflection fraction, and iron abundance were
fixed to 200~keV, 1, and $1 \Zsun$, respectively.

The model was fit to data 
between 2--10~keV from all four detectors. An acceptable fit was obtained with 
$\chi^2 = 1431.6 / 1423$ \textit{dof}, representing an improvement of 
$\Delta \chi^2 = 16.3$ with three additional parameters over {\tt pexrav} plus
cold absorption alone, and an improvement of
$\Delta \chi^2 = 30.2$ with three additional parameters over a power law plus
cold absorption alone.  The best-fit absorption column density was 
$1.32^{+0.08}_{-0.18} \times 10^{22}$ cm$^{-2}$, while the best-fit photon index was
$\Gamma = 1.46^{+0.02}_{-0.06}$. The iron-line parameters were well-constrained:
$\theta_0 = 49^{+8}_{-4}$ degrees, $\alpha = -2.7^{+0.8}_{-1.0}$, 
$F_{{\rm K}\alpha} = 1.4^{+0.3}_{-0.4}\times 10^{-5}$ 
\phpspsqcm~(corresponding to an equivalent width of $310^{+70}_{-90}\, \eV$).

\begin{figure}
\centerline{\psfig{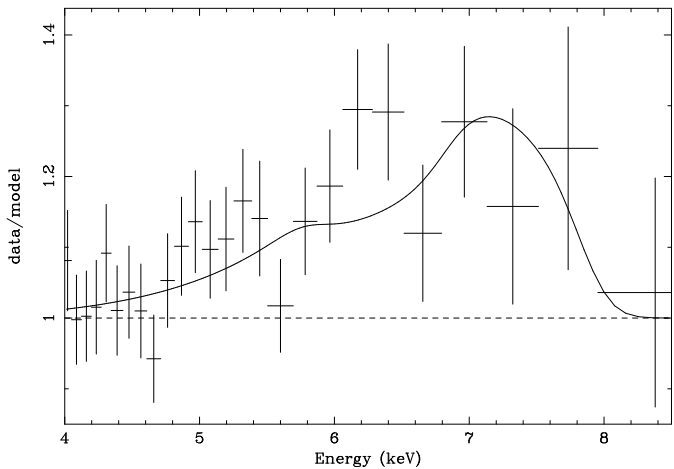}}
\caption{The ratio of the SIS spectrum to the best-fit continuum model. The
{\tt diskline} model is shown in the solid line. The 
excess signifies the clear presence of a broad iron K$\alpha$ line and is
described well by the {\tt diskline} model.}
\label{fig_ngc4395_fkaratio}
\end{figure}

We estimate the detection of a broad iron line in NGC 4395 to be significant 
at the $99\%$ level based on an F-test using a single-gaussian fit to the iron line.
The free parameters of the single-gaussian fit were the line energy $E_0$ and
the width $\sigma$, as well as the overall normalization. The line width was either 
frozen at $\sigma \equiv 0$ for a narrow line, 
or allowed to vary for a broad line.
The result of the fits was $\chi^2 = 1436 / 1424$ \textit{dof} vs.\ 
$\chi^2 = 1430 / 1423$ \textit{dof} for a narrow line vs.\ a broad line.
The best fit line parameters were $E_0 = 6.32^{+0.07}_{-0.03}$~keV for the 
narrow line and $E_0 = 6.26 \pm 0.20$~keV,
$\sigma=0.89^{+0.31}_{-0.21}$~eV for the broad line.
This significant detection of a broad iron line represents an
improvement over the marginal detection in the 
previous \textit{ASCA} observation of Iwasawa et al.\ (2000), thanks to the 
much larger signal-to-noise of this observation. The detected iron line is
plotted in Fig.\ \ref{fig_ngc4395_fkaratio} as the ratio of the SIS spectrum to 
the best-fit continuum model. The strength and broad width of the line is 
evident.

The spectrum below 4~keV was modeled using a {\sc cloudy} v90.04 table model, 
which describes one-zone, equilibrium, photoionized absorption \cite{Ferland91}. As mentioned
in Section \ref{sec_obs_and_datared}, small-aperture spectra were used 
for the analysis of the warm absorber, so as to minimize contamination
due to nearby circumnuclear sources. In order to remain consistent with the 
continuum fit above 2~keV, the {\sc cloudy} photon index was frozen at $\Gamma = 1.46$. The free
parameters of the fit were thus the column density and ionization parameter of the warm
absorber. The overall $\chi^2$ of the fit was acceptable (256.4/243 degrees of freedom),
but obvious trough-like residuals around 1.5--2~keV strongly suggested a more complex
pattern of absorption not well-accounted for by {\sc cloudy} alone. Also, we will see in Section 
\ref{sec_multizone_warmabs} that the single-zone {\sc cloudy} model fails to adequately describe
the flux-binned spectra. In particular, the pattern of absorption in the low-flux state cannot
be modelled using our simple single-zone {\sc cloudy} model.

The addition of two absorption edges (the multiplicative {\tt edge} model in {\sc xspec}) did the trick: 
the residuals were smoothed out, and the individual $\chi^2$s in the fit to the flux-binned 
spectra were all good (see Section \ref{sec_multizone_warmabs}).
The best-fit parameters for the {\sc cloudy} plus edges model are shown in the first 
row of Table \ref{tab_cloudy_edge_par}. We note that the two extra absorption edges are
fairly deep -- compare, for instance, with the results of 
Otani et al.\ (1996) for MCG--6-30-15. Note also that the inclusion of the two edges is
strongly favored by the F-test: a $\chi^2$ of 229.9/239 \textit{dof} with the edges,
versus a $\chi^2$ of 256.4/243 \textit{dof} corresponds to an F-test confidence level $> 99\%$.
The depth of these edges and the F-test both strongly suggest
a more complex picture of ionized absorption than that of the simple 
single-zone {\sc cloudy} model.

\begin{figure*}
\centerline{\psfig{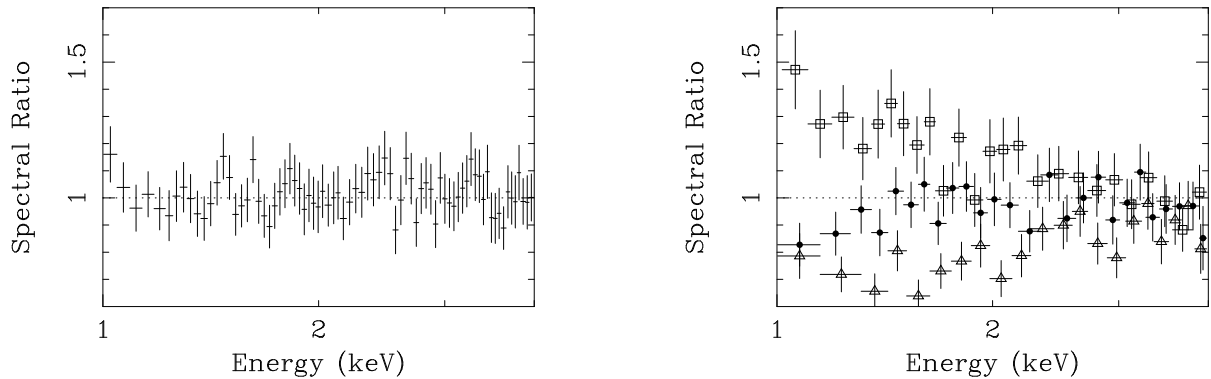}}
\caption{Ratios of the time-averaged spectrum (left panel) and the 
three flux-binned spectra (right panel) with the time-averaged best-fit model. Open triangles, 
solid circles, and open squares correspond to flux bins 1--3, respectively. The flux-binned 
spectra have been renormalized so that the 3--10~keV count-rates agree with the time-averaged 
spectrum. The absorption clearly lessens as the observed count rate increases. The largest 
change occurs in the 1--2~keV band, as is expected from the RMS variance spectrum shown 
in Fig.\ \ref{fig_rms}.}
\label{fig_spec_ratio}
\end{figure*}

\renewcommand{\arraystretch}{1.2}
\setlength{\tabcolsep}{1mm}
\begin{table*}
\begin{tabular}{|c|c|c|c|c|c|c|c|c|}
\hline
Bin & $\Gamma$ & $ N_{\rm W} $ & $\xi$ & $E_1$ & $\tau_1$ & $E_2$ & $\tau_2$ & $\chi^2$/\textit{dof} \\
         &          & $10^{22}$ \pcmsq  & \ergcmps & $\keV$ &  & $\keV$&          & \\[5pt]
\hline
\hline
0 & $1.46$ & $2.51^{+0.26}_{-0.28}$ & $203^{+60}_{-56}$ & $1.614^{+0.062}_{-0.062}$ & $0.153^{+0.058}_{-0.060}$
  & $1.890^{+0.073}_{-0.067}$ & $0.134^{+0.057}_{-0.061}$ & $229.9/239$ \\
\hline
1 & $1.46$    & $2.24^{+0.34}_{-0.22}$  & $121^{+47}_{-21}$   & $1.405^{+0.058}_{-0.053}$ & $0.429^{+0.112}_{-0.093}$ 
              & $1.857^{+0.066}_{-0.053}$  & $0.250^{+0.093}_{-0.088}$ & $94.6/85$ \\ 
2 &  ---      & ---                     & $133^{+53}_{-23}$   & ---                       & $0.009^{+0.092}_{-0.009}$ 
              & ---                        & $0.235^{+0.077}_{-0.074}$ & $95.1/101$ \\
3 &  ---      & ---                     & $220^{+130}_{-49}$   & ---                      & $< 0.060$ 
              & ---                        & $0.168^{+0.078}_{-0.091}$ & $81.2/94$ \\
\hline
\end{tabular}
\caption{The best-fit parameters for the {\sc cloudy} plus two edge model fitted to the 1--4~keV spectrum in each flux bin. 
Flux bin 0 corresponds to the time-averaged spectrum. Errors here and elsewhere are 1-$\sigma$ unless otherwise stated.
The dashes indicate that the corresponding parameter has been tied between flux bins.}
\label{tab_cloudy_edge_par}
\end{table*}

A promising alternative is the \textit{multi-zone} warm absorber consisting of two or more 
partially-ionized, absorbing regions with different physical 
properties and lying at different positions
along the line of sight. Spectral variability studies have strongly
implied the existence of a multi-zone warm absorber in several sources,
including MCG--6-30-15 \cite{Otani96,Morales2000}, NGC 3516 \cite{Kriss96}, 
and recently NGC 4051 \cite{Collinge2001}. In the previous short \textit{ASCA}
observation of NGC 4395, a two-zone warm absorber (one constant and less ionized,
the other variable and more ionized) was argued for based on 
an unphysical \textit{decrease} in column density and ionization parameter 
observed during a flare \cite{Iwasawa2000}. 

The result of a two-zone warm absorber fit to the time-averaged spectrum (below 4~keV) is shown
in the first row (flux bin 0) of Table \ref{tab_cloudy_absori_par}. The model used was a hybrid 
{\sc cloudy}/{\tt absori} model, with the former (latter) used to model the 
constant (variable) component. The ionization parameters and column densities of the two models,
along with the underlying photon index, were left free to vary. The temperature of the
accretion disk in {\tt absori} was fixed to $10^6$~K. The model clearly describes the 
time-averaged spectrum well. The column density and ionization parameter of the constant component
agrees well with the {\sc cloudy} plus edges fit described previously (see the 
first row of Table \ref{tab_cloudy_edge_par}). This gives us confidence in the validity of the arguments
above for the necessity of a more complex, multi-zone warm-absorber model. 
As another consistency check, the column densities of the two zones are in rather good agreement with those
obtained in the previous observation \cite{Iwasawa2000}. Meanwhile the ionization parameters,
especially of the variable component, are considerably higher than before. 

In their recent analysis of a 17~ks \textit{Chandra} observation of NGC 4395, Moran et al. (2001) also 
find that a two-zone warm absorber model can provide a satisfactory fit to the time-averaged data. 
However they fix the underlying photon index to $\Gamma = 0.56$, 
based on a previous \textit{single-zone} warm absorber fit. In our analysis, with our 
much larger statistical sample, we are able to fit independently for the photon index in the 2--10~keV
range with a more sophisticated model consisting of a power-law, reflection, and broad iron line. As
a result, we find a much steeper, less exotic photon index of $\Gamma = 1.46$. 
We are therefore led to a conclusion quite different from that of Moran et al. (2001):
NGC 4395 possesses an intrinsically steep power-law continuum with 
$\Gamma = 1.46$, which is subsequently flattened and modified by multi-zone, ionized absorption, 
especially around 1.5--2~keV.

\renewcommand{\arraystretch}{1.2}
\setlength{\tabcolsep}{1mm}
\begin{table*}
\begin{tabular}{|c|c|c|c|c|c|c|}
\hline
Bin & $\Gamma$ & $N^{\rm c}_{\rm W}$   & $\xi^{\rm c}$ & $N^{\rm v}_{\rm W}$  & $\xi^{\rm v}$ & $\chi^2$/\textit{dof} \\
    &          & $10^{22}$ \pcmsq  & \ergcmps  & $10^{22}$ \pcmsq & \ergcmps  &                       \\[5pt]
\hline
\hline
0 & $1.46$ & $2.45^{+0.25}_{-0.32}$ & $200^{+55}_{-60}$ & $7.99^{+3.33}_{-2.45}$ & $481^{+123}_{-111}$ & $232.8/241$ \\
\hline
1 & $1.46$ & $1.91^{+0.31}_{-0.21}$ & $106^{+26}_{-8}$  & $5.80^{+0.63}_{-1.23}$ & $157^{+15}_{-13}$   & $99.2/87$ \\
2 & ---    & ---                    & $103^{+16}_{-4}$  & ---                    & $436^{+100}_{-83}$  & $96.5/102$ \\
3 & ---    & ---                    & $159^{+27}_{-16}$ & ---                    & $461^{+85}_{-88}$ & $80.6/95$ \\
\hline
\end{tabular}
\caption{The best-fit parameters for the two-zone {\sc cloudy} plus {\tt absori} model.
The ``c'' and ``v'' superscripts denote the constant and the variable warm absorber zones, 
respectively. The dashes indicate that the corresponding parameter has been tied between flux bins.}
\label{tab_cloudy_absori_par}
\end{table*}

A more revealing test of the two-zone warm absorber model is whether it can describe the response
of the spectrum to changes in source flux. In the next section, we shall 
use time-resolved spectral analysis to investigate whether 
the multi-zone model proposed previously for NGC 4395 can still provide
an adequate description of the data, given the higher signal-to-noise and
longer exposure time of this observation. 

\subsection{Time-Resolved Analysis: Multizone Warm Absorber}
\label{sec_multizone_warmabs}

To investigate the spectral variability of the warm absorber, 
the data were binned with respect to $F_{1.2-5}$, the SIS0 count rate 
in the 1.2--5~keV band. Three bins were used, chosen to contain approximately 
the same number of counts, and defined to be $F_{1.2-5} < 0.025$ cts/s, 
$0.025 \le F_{1.2-5} < 0.04$ cts/s, and $F_{1.2-5} \ge 0.04$ cts/s. The bins 
are numbered 1--3 in order of increasing flux; flux bin 0 denotes the time-averaged spectrum.

The spectral ratios between the data in flux bins 0--3 and the best-fit time-averaged {\sc cloudy} plus
edges model are plotted in Fig.\ \ref{fig_spec_ratio}. 
The model clearly gives a reasonable fit to the time-averaged data, although the residuals 
still suggest the presence of a complex of weak absorption edges 
(and possibly emission lines) that our simple two extra edges fail to take into account. 
In flux bins 1--3, the 3--10~keV count rates have been renormalized to that of the 
time-averaged spectrum. Since the 3--10~keV 
count rate should be largely unaffected by absorption, the renormalized ratios plotted
here should give at least an approximate idea of the variation in absorption. 
From the plots, it is clear that the warm absorber is changing dramatically with flux.
In the low flux bin (bin 1), the spectral ratio reveals a shape characteristic of one or more
absorption edges. As we increase in flux, the warm absorber appears to decrease in
strength, suggesting that the elements responsible for it are being pushed to higher ionization
states due to the increased amount of incident radiation.

We can confirm this interpretation by fitting the {\sc cloudy} plus edge model to the flux-binned 
spectra. The photon index, being poorly-constrained,
was fixed to the best-fit value of the time-averaged 2--10~keV spectrum. The {\sc cloudy} column density
and the threshold energies of the edges were tied between the flux-binned spectra. 
The result of the fits are shown in rows 2--4 of Table \ref{tab_cloudy_edge_par}. The $\chi^2$
is clearly acceptable in the individual flux bins.
Also, as mentioned in Section \ref{sec_tavg}, the two additional edges are necessary in order to 
obtain a good fit in the first flux bin. Redoing the fit with {\sc cloudy} alone results
in $\chi^2$s of 118.8/89, 107.3/103, and 85.8/96 \textit{dof} in flux bins 1-3, respectively.
Comparing with the $\chi^2$ column in Table \ref{tab_cloudy_edge_par}, we see that two additional edges
have a very significant impact on the low-flux bin.
The fits confirm the existence of a constant underlying warm-absorber component -- while
intriguing, the increase in ionization parameter $\xi$ from the second to the
third flux bin is not statistically significant. The spectral variation across flux 
bins observed in Fig.\ \ref{fig_spec_ratio} is primarily accounted for by changes in the
optical depths of the photoionization edges. 
We therefore make the obvious identification of the constant and variable warm absorber
with the parameters of {\sc cloudy} and of the extra edges, respectively. 

The two-zone warm absorber model confirms this identification. Fits using the {\sc cloudy}/{\tt absori}
model described in Section \ref{sec_tavg} were performed with the column densities of the
constant and variable components frozen tied between the flux bins. The results
are shown in rows 2--4 of Table \ref{tab_cloudy_absori_par}. The two-zone model
describes well the spectral variation with flux, and the variable component behaves in the expected manner,
increasing in ionization parameter with increasing source flux. 
There is strong evidence, therefore, in favor of the two-zone warm absorber model 
proposed for NGC 4395 by Iwasawa et al.\ (2000).
Absorption around 1.5--2~keV implies that the variable zone is, on average,
highly ionized. Depending on the ionization state of the zone, the dominant absorption at
these energies should be due to He/H-like Mg and Si. We should note that the high ionization of
the variable absorber in all flux states provides an \textit{a posteriori} justification for our model
consisting of simple multiplication of the two absorbers ({\sc cloudy} $\times$ {\tt absori}). This
is fortunate, since the proper approach -- in which the output from the inner absorber is used 
as the input for the outer one -- would have proved much more difficult.

Perhaps it is also worthwhile to note that the slight increase
in the ionization parameter of the constant component ($\xi_c$) from the second to the third bin persists
in the {\sc cloudy}/{\tt absori} fit. But again
it is not statistically significant. In spite of this, it is still very interesting
that the ionization parameter of the variable component ($\xi_v$) 
seems to level off in the high-flux state. Combining this
fact with the slight increase in $\xi_c$ leads us to suggest the following interpretation
of the flux-binned fit: the variable component, assumed to be closer to the
nuclear source, becomes completely ionized, and therefore transparent, in the high-flux state. As a
result, the constant outer absorber begins to be ionized, and its ionization parameter increases.

A more detailed interpretation of our results and a more general discussion of the warm absorber
in NGC 4395 would be too extensive to include in this paper. We should point out that some of 
the properties of a warm absorber in a small black hole, low luminosity system such as NGC4395 are expected
to differ from those in ordinary Seyferts. For instance, the variability
time scale of a warm absorber around a normal Seyfert is typically months to years, if the ionized gas
is coming from the inner edge of a torus around 1~\pc~\cite{Kriss2001}. On the other hand, $10^3$-s variability
in the warm absorber of NGC 4395 is entirely possible, given the low luminosity and high ionization
parameter of this source. The properties of warm absorbers in low-luminosity systems such as NGC 4395
is an interesting topic in its own right, and we postpone a more complete discussion 
to a future publication.

\section{Discussion: Estimating the Black Hole Mass}
\label{sec_bhmass}

The break in the power spectrum from $\alpha \sim 1$ to $\alpha \sim 1.8$ at $3.2 \times 10^{-4} \Hz$ found in Section \ref{sec_PDS} 
gives us a characteristic variability timescale for NGC 4395. Assuming that the variability timescale of a 
source scales roughly with its size
(i.e.\ a larger source has a longer variability timescale), we can use the measured break frequency and the PDS of a more 
well-studied source such as Cyg X-1 to obtain an estimate of the black hole mass of NGC 4395. This method has been previously applied
to several AGN (e.g.\ Nowak et al.\ 2000, Hayashida et al.\ 1998, Iwasawa et al.\ 2000), and tends to give mass 
estimates lower (by as much as 1 or 2 orders of magnitude in some cases) than those 
obtained from other methods such as X-ray variability, emission line broadening, and 
modeling the big blue bump \cite{Hayashida98}. 

The $\alpha \sim 1$ to $\alpha \sim 1.8$ break frequency of Cyg X-1 is known to 
fluctuate between 1--10 \Hz~\cite{Belloni90} over time.
The mass of Cyg X-1 is estimated to be $\sim 10$ \Msun~\cite{Herrero95}. Therefore, if the break frequency scales with the size
of the system, we estimate the mass of NGC 4395 to be $M_{\rm BH} \sim 10^4-10^5$ \Msun. This is at 
the high end of the hard upper limit of 
$M_{\rm BH} \approxlt 8\times 10^5 \Msun$ 
obtained in recent kinematical studies by Filippenko et al. (2002); but
it is encouraging that it is in complete agreement 
with their ``most probable'' range. 
We can also compare the PDS of NGC 4395 with those of other AGN which have independent, accurate mass measurements. 
Perhaps the best example of an AGN with a well-determined power spectrum and mass is NGC 5548.
Reverberation mapping and kinematical studies of NGC 5548 give a central black hole mass of $(6.8 \pm 2.1) \times 10^7 \Msun$ 
\cite{Peterson99}, and the PDS agrees with that of Cyg X-1 with a 
frequency scale factor of $\sim 10^{5}$--$10^{6}$ \cite{Chiang2000}. 
Using NGC 5548 as a template would therefore imply $M_{\rm BH} \sim 10^5$ $M_{\odot}$ in NGC 4395, 
in agreement (albeit at the high end) with the estimate using Cyg X-1. 

NGC 4395 is a bulgeless dwarf galaxy (or with a very faint bulge of
late spiral).  Based on a study of M33, a nearby bulgeless galaxy,
Gebhardt et al. (2001) suggest that no supermassive black hole is
present in a galaxy with no bulge. They pointed out that a black hole
of 100~$\Msun$ is sufficient for the luminosity of the active nucleus
of NGC 4395 if it is operating at the Eddington limit. The bulge
luminosity of NGC 4395 is indeed very low ($M_{\rm B,bulge}=-10.36$,
Ho et al 1997), similar to that of the nucleus of M33, indicating that
the black hole driving the active nucleus of NGC 4395 could be as
small as 100~$\Msun$. However, while there is an order of magnitude
uncertainty in our estimate, the X-ray variability results appear to
favor the black hole mass to be much larger.  The estimates using the
optical emission-line properties have also given similar results (Lira
et al 1999; Kraemer et al 1999). We also point out that considerable
obscuration to the nucleus of NGC 4395, as revealed by hard X-ray
observations, leads to the true (optical) nuclear luminosity being
underestimated. With a modest bolometric correction ($L_{\rm
  bol}/L_{\rm 2-10keV}=10$), the mean absorption-corrected 2--10 keV
  luminosity of $4\times 10^{39}$\ergps\ requires a black hole of
  $\simeq 300 M_{\odot}$ operating at the Eddington limit. The
  observed light curve shows that the X-ray flux occasionally reaches
  6 times higher than the mean value, which implies an even heavier
  black hole of $\simeq 2\times 10^3 M_{\odot}$. Moreover, the fact
  that Seyfert galaxies usually operate at a fraction (a few per cent)
  of their Eddington luminosity points to that the black hole in
  NGC4395 might in fact be as large as $10^4$--$10^5 \Msun$.
  Therefore, either NGC 4395 might be peculiar in that it has a
  supermassive black hole at the center despite the lack of a
  significant bulge, or the X-ray source in NGC 4395 exhibits
  variability in a way which apparently does not scale by the black
  hole mass.

\section*{Acknowledgments}
We thank Simon Vaughan for insights into the subtleties of PDS and timing analysis, Raquel Morales for useful discussions regarding the multi-zone warm absorber and Luis Ho for information on thier unpublished result on the black hole mass of NGC 4395. For support, we thank the 
Herchel Smith Fellowship (DCS), PPARC (KI), and the Royal Society (ACF).

\end{document}